\documentclass[runningheads]{llncs}

\usepackage{graphicx}
\usepackage{mathpartir}
\usepackage{booktabs}   %
\usepackage{subcaption} %
\usepackage{formal-grammar}
\usepackage{caption}
\usepackage{subcaption}
\usepackage{pgfplotstable}

\usepackage{multirow}
\usepackage{graphics}

\usepackage{cite}
\usepackage{tabularx}
\usepackage{hyperref}
\usepackage{amsfonts}
\usepackage{ifthen}
\usepackage{fontawesome5}
\usepackage{tikz}
\usetikzlibrary{fit,arrows,calc,positioning}

\usepackage{xcolor}
\usepackage{listings}
\definecolor{GrayCodeBlock}{RGB}{241,241,241}
\definecolor{BlackText}{RGB}{110,107,94}
\definecolor{RedTypename}{RGB}{182,86,17}
\definecolor{GreenString}{RGB}{96,172,57}
\definecolor{PurpleKeyword}{RGB}{184,84,212}
\definecolor{GrayComment}{RGB}{170,170,170}
\definecolor{GoldDocumentation}{RGB}{180,165,45}
\definecolor{codegray}{rgb}{0.5,0.5,0.5}
\definecolor{backcolour}{rgb}{0.95,0.95,0.92}
\lstdefinelanguage{rust}
{
    breakatwhitespace=false,
    breaklines=true, 
    keepspaces=true,                 
    numbers=left,                    
    numbersep=5pt,                  
    showspaces=false,                
    showstringspaces=false,
    showtabs=false,                  
    tabsize=2,
    columns=fullflexible,
    keepspaces=true,
    frame=single,
    framesep=0pt,
    framerule=0pt,
    framexleftmargin=4pt,
    framexrightmargin=4pt,
    framextopmargin=5pt,
    framexbottommargin=3pt,
    xleftmargin=4pt,
    xrightmargin=4pt,
    basicstyle=\ttfamily,%
    keywords={
        typedef,
        true,false,
        unsafe,async,await,move,
        use,pub,crate,super,self,mod,
        struct,enum,fn,const,static,let,mut,ref,type,impl,dyn,trait,where,as,
        break,continue,if,else,while,for,loop,match,return,yield,in,NULL
    },
    keywordstyle=\color{magenta},%
    numberstyle=\tiny\color{codegray},
    ndkeywords={
        bool,u8,u16,u32,u64,u128,i8,i16,i32,i64,i128,char,str,
        Self,Option,Some,None,Result,Ok,Err,String,Box,Vec,Rc,Arc,Cell,RefCell,HashMap,BTreeMap,
        macro_rules
    },
    ndkeywordstyle=\color{RedTypename},
    comment=[l][\color{gray}\slshape]{//},
    morecomment=[s][\color{gray}\slshape]{/*}{*/},
    morecomment=[l][\color{GoldDocumentation}\slshape]{///},
    morecomment=[s][\color{GoldDocumentation}\slshape]{/*!}{*/},
    morecomment=[l][\color{GoldDocumentation}\slshape]{//!},
    morecomment=[s][\color{RedTypename}]{\#![}{]},
    morecomment=[s][\color{RedTypename}]{\#[}{]},
    stringstyle=\color{GreenString},
    string=[b]"
}
\usepackage{amsmath} %
\usepackage{algpseudocode} %
\usepackage{xspace} 
\usepackage[ruled,vlined]{algorithm2e}

\input{macro}

\begin{document}

\title{Ownership guided C to Rust translation} 

\author{Hanliang Zhang\inst{1}\and Cristina David\inst{1}\and Yijun Yu\inst{2} \and Meng Wang\inst{1}} 
\institute{University of Bristol\\
\email{\{pd21541,cristina.david,meng.wang\}@bristol.ac.uk}
\and
The Open University\\
\email{yijun.yu@open.ac.uk}}
\maketitle              %

\begin{abstract}
Dubbed a safer C, Rust is a modern programming language that combines memory safety and low-level control. 
This interesting combination has made Rust very popular among developers and there is a
growing trend of migrating legacy codebases (very often in C) to Rust. In this paper, we present a 
C to Rust translation approach centred around static ownership analysis. We design a suite of 
analyses that infer ownership models of C pointers and automatically translate the pointers into safe Rust equivalents.  
The resulting tool, \name, scales to real-world codebases (half a million lines of code in less than 10 seconds) and 
achieves a high conversion rate.  
\end{abstract}

\section{Introduction} \label{sec:intro}

Rust~\cite{matsakis2014rust} is a modern programming language which features an exciting combination of
memory safety and low-level control. 
In particular, Rust takes inspiration from ownership and substructural (mostly affine and linear) types and 
 restricts certain memory accesses to their \emph{owners}. This means  that 
 although multiple pointers to the resource can co-exist (through a mechanism known as \emph{borrowing}), 
 certain operations can only be performed by the owner.
The Rust compiler is able to statically verify the ownership constraints and consequently guarantee memory and thread safety.
This distinctive advantage of provable safety makes Rust a very popular language, and the prospect of migrating 
legacy codebases in C to Rust is very appealing.

In response to this demand, automated tools translating C code to Rust emerge from 
both industry and academia~\cite{c2rust,DBLP:journals/pacmpl/EmreSDH21,CRustS}.
Among them, the industrial strength translator C2Rust~\cite{c2rust} rewrites C code into the Rust syntax 
while preserving the original semantics. The translation does not synthesise an ownership model and 
thus is not able to do more than replicating the unsafe use of pointers in C. And consequently, the Rust code must be labelled 
with the \lstinline{unsafe} keyword which allows certain actions that are not checked by the compiler. %
More recent work focuses on reducing this unsafe labelling. In particular, the tool Laertes~\cite{DBLP:journals/pacmpl/EmreSDH21} aims to 
rewrite the (unsafe) code produced by C2Rust by searching the solution space guided by the type error messages from the Rust compiler. 
This is ground breaking, as for the first time proper Rust code beyond a line-by-line direct 
conversion from the original C source may be synthesised. 
On the other hand, the limit of the trial-and-error approach is also clear: the system does not
support the reasoning of the generation process, nor create any new understanding of the target code (other than the obvious fact that it
compiles successfully).

In this paper, we take a more principled approach by developing a novel ownership analysis of pointers that is efficient (scaling to large programs (half a million LOC in less than 10 seconds)),
sophisticated (handling nested pointers and inductively-defined data structures), and precise (being field and flow sensitive).
Our ownership analysis is both scalable and precise owing to a strengthening assumption we make about the Rust ownership model, which obviates
the need for an aliasing analysis.

The primary goal of this analysis is of course to facilitate C to Rust translation. Indeed, as we will see 
in the rest of the paper, an automated translation system is built to encode the ownership models in the generated Rust code which is then proven safe by the Rust compiler. 
But there is more. In contrast to trying the Rust compiler as common in existing approaches~\cite{DBLP:journals/pacmpl/EmreSDH21,CRustS}, this analysis approach 
actually extracts new knowledge of ownership from code, which may lead to a range of utilities including preventing memory leaks, identifying 
inherently unsafe code fragments, detecting memory bugs, and so on. Specifically, in this paper we 
\begin{itemize}
\item design a scalable and precise ownership analysis 
  that is able to handle complex inductively-defined data structures and nested pointers. (Section~\ref{sec:ownership})
\item develop a refactoring mechanism for Rust leveraging ownership analyses to enhance code safety. (Section~\ref{sec:translation})
\item implement a prototype tool (\name, standing for C to Rust OWNership guided translation) that translates C code into Rust with enhanced safety. (Section~\ref{sec:discussion})
\item evaluate \name with a benchmark suite including commonly used data structure libraries and real-world projects (ranging from 150 to half a million LOC) and compare the result with the state-of-the-art. (Section~\ref{sec:experiments})
\end{itemize}

\section{Background} \label{sec:background}
We start by giving a brief introduction of Rust, in particular its ownership system and the use of pointers, as they
are core to memory safely.

\subsection{Rust ownership model} \label{sec:rustownership}

Ownership in Rust denotes a set of rules that govern how a Rust program manages memory~\cite{matsakis2014rust}.
The idea is to associate each value with a \emph{unique} owner. This feature is useful for memory management. 
For example, when the owner goes out of scope, the memory
allocated for the value can be automatically recycled. 
\begin{lstlisting}
let mut v = ...
let mut u = v;         // ownership is transferred to u
\end{lstlisting}
In the above snippet, the assignment of \lstinline{v} to \lstinline{u} also transfers ownership, after which
\lstinline{v} is dropped from the scope and cannot be used again.

This permanent transfer of ownership gives strong guarantees but can be cumbersome to manage in programming. 
For temporal transfer of ownership (known as \emph{borrowing}), one can use 
a \emph{reference} (marked by an ampersand). For example, in \lstinline{f(&mut x)}, 
the ownership of \lstinline{x}'s value is temporally borrowed by to the function call until its return.

This concept of time creates another dimension of ownership management known as \emph{lifetime}. 
For mutable references (as marked by \lstinline{mut} in the above examples), the rule is relatively simple:
only one mutable reference is allowed at anytime. But for immutable references (the ones without the \lstinline{mut} marking),
multiple of them can coexist as long as there isn't any mutable reference at the same time. 
As one can expect, this interaction of mutable and immutable references, and their lifetimes is highly
non-trivial. In this paper, we focus on analysing mutable references as they are far more common than immutable ones.

\subsection{Pointer types in Rust} \label{sec:backgroud-pointers}
Rust has a richer pointer system than C.
The primitive C-style pointers (written as \lstinline{*const T} or \lstinline{*mut T}) are known as \emph{raw pointers}, which 
are ignored by the Rust compiler for ownership and lifetime checks. Raw pointers are a major
source of unsafe Rust (more below). Idiomatic Rust instead advocates \emph{box pointers} (written as \lstinline{Box<T>}) as owning pointers that uniquely own heap allocations, as well as \emph{references} (written as \lstinline{&mut T} or \lstinline{& T} as discussed in the 
previous subsection) as non-owning pointers that are used to access values owned by others.

C-style array pointers are represented in Rust as references to arrays and slice references, with array bounds
known at compile time and runtime, respectively. The creation of meta-data such as array bounds is  
beyond the scope of ownership analysis. In this work, we keep array pointers as raw pointers in the translated code.

\subsection{Unsafe Rust}  
As a pragmatic design, Rust allows programs to contain features that cannot be verified 
by the compiler as memory safe. This includes dereferencing raw pointers, calling low level
 functions, and so on. Such uses must to marked by the \lstinline{unsafe} keyword and form
fragments of \emph{unsafe Rust}. It is worth noting that \lstinline{unsafe}
does not turn off compiler checks; safe pointers are still checked. 

The main purpose of unsafe Rust is to support low-level systems programming. But it can 
also be used for other reasons. 
For example, \crust~\cite{c2rust} directly translates C pointers into raw pointers. Without
unsafe Rust, the generated code would not compile.

\section{Overview} \label{sec:motivation}
In this section, we present an overview of \name via two examples. The first example provides a detailed description of the \lstinline{push} method
for a singly-linked list, whereas the second shows a snippet from a real-world benchmark.

\begin{figure}
    \centering
        \begin{subfigure}[t]{0.3\textwidth}
            \begin{lstlisting}[language=c, numbers=left, firstnumber=1, mathescape=true, basicstyle=\tiny\ttfamily]{Name}
struct Node {
    int data;
    struct Node * next;
};
    
struct List {
    Node * head; 
};

void push(struct List* list, int new_data) {
    struct Node* new_node = (struct Node*) malloc(sizeof(struct Node));
    new_node->data  = new_data;
    new_node->next = list->head;
    list->head    = new_node;
}
            \end{lstlisting}
            \caption{C code}
            \label{fig:llist-constr-c}
        \end{subfigure}
    \hfill
    \begin{subfigure}[t]{0.3\textwidth} %
      \begin{lstlisting}[language=Rust, numbers=left, firstnumber=1, mathescape=true, basicstyle=\tiny\ttfamily]{Name}
#[repr(C)]
#[derive(Copy, Clone)]
pub struct Node {
    pub data: i32,
    pub next: *mut Node,
}
        
#[repr(C)]
#[derive(Copy, Clone)]      
pub struct List {
    pub head: *mut Node,
} 
      
pub unsafe extern "C" fn push(mut list: *mut List, mut new_data: i32) {
    let mut new_node = malloc(::std::mem::size_of::<Node>() as libc::c_ulong) as *mut Node;
    (*new_node).data = new_data;
    (*new_node).next = (*list).head;
    (*list).head = new_node;
}
        \end{lstlisting}
        \caption{\crust result}
        \label{fig:llist-constr-rust}
        \end{subfigure}
        \hfill
        \begin{subfigure}[t]{0.3\textwidth}
        \begin{lstlisting}[language=Rust, numbers=left, firstnumber=1, mathescape=true, basicstyle=\tiny\ttfamily]{Name}
#[repr(C)]
pub struct Node {
    pub data: i32,
    pub next: Option<Box<Node>>,
}
        
#[repr(C)]   
pub struct List {
    pub head: Option<Box<Node>>,
}

pub unsafe extern "C" fn push(mut list: Option<&mut List>, mut new_data: i32) {
    let mut new_node = Some(Box::new(<Node as Default>::default()));
    (*new_node.as_deref_mut().unwrap()).data = new_data;
    (*new_node.as_deref_mut().unwrap()).next = (*list.as_deref_mut().unwrap()).head.take();
    (*list.as_deref_mut().unwrap()).head = new_node;
}
        \end{lstlisting}
        \caption{\name result}
        \label{fig:llist-constr-crown}
        \end{subfigure}
        \caption{Pushing into a singly-linked list}
        \label{fig:llist-constr}

\end{figure}

\subsection{Pushing into a singly-linked list}

The C code of function \lstinline{push} in figure~\ref{fig:llist-constr-c} allocates a new node where it stores the data received as argument.  The new node subsequently becomes the head of \lstinline{list}. This code is translated by \lstinline{c2rust} to the Rust code in figure~\ref{fig:llist-constr-rust}. Notably, the \lstinline{c2rust} translation is syntax-based and
simply changes all the C pointers to \lstinline{*mut} raw pointers. Given that
dereferencing raw pointers is considered an unsafe operation in Rust (e.g. the dereferencing of \lstinline{new_node} at line 16 in figure~\ref{fig:llist-constr-rust}), method
\lstinline{push} must be annotated with the \lstinline{unsafe} keyword.
Additionally, \lstinline{c2rust} introduces two directives for the two struct definitions, 
\lstinline{#[repr(C)]} and \lstinline{#[derive(Copy, Clone)]}. The former keeps the data
layout the same as in C for possible interoperation, and the latter instructs that the corresponding type can only be duplicated through copying.

While \lstinline{c2rust} uses raw pointers in the translation, the ownership scheme in figure~\ref{fig:llist-constr-rust} obeys the Rust ownership model, meaning that
the raw pointers could be translated to safe ones.
A pointer to a newly allocated node is assigned to \lstinline{new_node} at line 15. This allows us to infer that the ownership of the newly allocated node belongs to \lstinline{new_node}.
Then, at line 18, the ownership is transferred from \lstinline{new_node} to \lstinline{(*list).head}.
Additionally, if \lstinline{(*list).head} owns any memory object prior to line 17, then its ownership is transferred to
\lstinline{(*new_node).next} at line 17.
This ownership scheme corresponds to safe pointer use:
(i) each memory object is associated with a unique owner and (ii) it is dropped 
when its owner goes out of scope. As an illustration for (i), when the ownership of the newly allocated memory is transferred from \lstinline{new_node} to  \lstinline{(*list).head} at line 18,  \lstinline{(*list).head} becomes the unique owner, whereas \lstinline{new_node} is made invalid and it is no longer used.
For (ii), given that argument \lstinline{list} of \lstinline{push} is an output parameter (i.e. a parameter that can be accessed from outside the function),
we assume that it must be owning on exit from the method. 
Thus, no memory object is dropped in the \lstinline{push} method, but rather returned to the caller.

\name infers the ownership information of the code translated by \lstinline{c2rust}, and uses it to translate the code to safer Rust in figure~\ref{fig:llist-constr-crown}.
As explained next, \name first retypes raw pointers into safe pointers based on the ownership information, and then rewrites their
uses.

{\bf Retyping pointers in \name.} %
If a pointer owns a memory object at \emph{any point within its scope},
\name retypes it into a \boxptr pointer. %
For instance, in figure~\ref{fig:llist-constr-crown}, local variable \lstinline{new_node} is retyped to be \optionboxptr{Node} (safe pointer types are wrapped into Option to account for null pointer values).
Variable \lstinline{new_node} is non-owning
upon function entry, becomes owning at line 13 and ownership is transferred out again at line 16. 

For struct fields, \name considers all the code in the scope of the struct declaration.
If a struct field owns a memory object at \emph{any point within the scope of its struct declaration}, then it is retyped to \lstinline{Box}.
In figure~\ref{fig:llist-constr-rust}, fields \lstinline{next} and \lstinline{head} are accessed via access paths \lstinline{(*new_node).next} and  \lstinline{(*list).head}, and given ownership at lines 17 and 18, respectively. Consequently, they are retyped to \lstinline{Box} at lines 4 and 9 in figure~\ref{fig:llist-constr-crown}, respectively.

A special case is that of output parameters, e.g. \lstinline{list} in our example. 
For such parameters, although they may be owning, %
\name retypes them to \lstinline{&mut} in order to enable borrowing.
In \lstinline{push}, the input argument \lstinline{list} is retyped
to \optionmutref{List}.

{\bf Rewriting pointer uses in \name.} After retyping pointers,
\name rewrites their uses. The rewrite process takes into consideration both their new type and the context in which they are being used.
Due to the Rust semantics, the rewrite rules are slightly intricate (see section~\ref{sec:translation}).
For instance, the dereference of \lstinline{new_node} at line 14 is rewritten
to \lstinline{(*new_node).as_deref_mut().unwrap()} as it needs to be mutated and the optional part of the \boxptr needs to be unwrapped.
Similarly, at line 15, \lstinline{(*list).head} is rewritten to be \lstinline{((*list.as_deref_mut()).unwrap()).head.take()} as the LHS
of the assignment expects a \boxptr pointer. %

\subsection{Freeing an argument list in bzip2} \label{sec:overview-loop}
We next show the transformation of a real-world code snippet with a loop structure:
a piece of code in \lstinline{bzip2} that frees argument lists.
\begin{figure}
    \centering
        \begin{subfigure}[t]{0.3\textwidth}
        \begin{lstlisting}[language=c, numbers=left, firstnumber=1, mathescape=true, basicstyle=\tiny\ttfamily]{Name}
typedef
struct zzzz {
    Char        *name;
    struct zzzz *link;
}
Cell;
[...]
Cell* aa = argList;
while (aa != NULL) {
    Cell* aa2 = aa->link;
    if (aa->name)
        free(aa->name);
    free(aa);
    aa = aa2;
}
        \end{lstlisting}
        \caption{C definition}
        \label{fig:cell-c}
        \end{subfigure}
    \hfill
        \begin{subfigure}[t]{0.3\textwidth}
        \begin{lstlisting}[language=Rust, numbers=left, firstnumber=1, mathescape=true, basicstyle=\tiny\ttfamily]{Name}
#[derive(Copy, Clone)]
#[repr(C)]
pub struct zzzz {
    pub name: *mut Char,
    pub link: *mut zzzz,
}
pub type Cell = zzzz;
[...]
let mut aa = argList;
while !aa.is_null() {
    let mut aa2 = (*aa).link;
    if !(*aa).name.is_null() {
        free((*aa).name as *mut libc::c_void);
    }
    free(aa as *mut libc::c_void);
    aa = aa2;
}
        \end{lstlisting}
        \caption{\crust result}
        \label{fig:cell-rust}
        \end{subfigure}
        \hfill
        \begin{subfigure}[t]{0.3\textwidth}
        \begin{lstlisting}[language=Rust, numbers=left, firstnumber=1, mathescape=true, basicstyle=\tiny\ttfamily]{Name}
#[repr(C)]
pub struct zzzz {
    pub name: *mut /* owning */ Char,
    pub link: Option<Box<zzzz>>,
}
pub type Cell = zzzz;
[...]
let mut aa = argList;
while !aa.as_deref().is_none() {
    let mut aa2 = (*aa.as_deref_mut().unwrap()).link.take();
    if !(*aa.as_deref().unwrap()).name.is_null {
        free((*aa.as_deref().unwrap()).name as *mut libc::c_void);
    }
    aa = aa2;
}
        \end{lstlisting}
        \caption{\name result}
        \label{fig:cell-crown}
        \end{subfigure}
        \caption{Freeing an argument list}
        \label{fig:cell-declaration}
\end{figure}
\lstinline{bzip2} defines a singly-linked list like structure, \lstinline{Cell}, that holds a list of
argument names. In figure~\ref{fig:cell-declaration}, we extract
from the source code a snippet that frees the argument lists.
Here, the local variable \lstinline{argList} is an already constructed argument list,
and \lstinline{Char} is a type alias to C-style characters.

\name accurately infers an ownership scheme for this snippet. Firstly, ownership of \lstinline{argList}
is transferred to \lstinline{aa}, which is to be freed in the subsequent loop. Inside the loop, ownership of
\lstinline{link} accessed from \lstinline{aa} is firstly transferred to \lstinline{aa2}, then
ownership of \lstinline{name} accessed from \lstinline{aa} is released in a call to \lstinline{free}.
After the loop, ownership of \lstinline{aa} is also released.
Last of all, \lstinline{aa} regains ownership from \lstinline{aa2}. 

{\bf Handling of loops.} For loops, \name only analyses their body once as that will already expose all the ownership information.
For inductively defined data structures such as \lstinline{Cell}, while further unrolling of loop bodies explores the data structures deeper, it does not expose
any new struct fields: pointer variables and pointer struct fields do not change ownership between loop iterations.
Additionally,
\name emits constraints that equate the ownership of all local pointers at 
the loop entry and exit. 
For example, the ownership statuses of \lstinline{aa} and \lstinline{aa2} 
at loop entry are made equal with those at loop exit, and inferred to be owning and non-owning, respectively.

{\bf Handling of null pointers.}
It is a common C idiom for pointers to be checked against null after malloc or before
free: \lstinline$if !p.is_null() { free(p); }$. This could be problematic since the then-branch and the else-branch
 would have conflicting ownership statuses for \lstinline{p}. We adopt a similar solution as 
\cite{DBLP:conf/pldi/HeineL03}: we insert an explicit null assignment in the null branch
\lstinline$if  !p.is_null()  { free(p); } else { p = ptr::null_mut(); }$.
As we treat null pointers
as both owning and non-owning, the ownership of  \lstinline{p}
will be dictated by the non-null branch, enabling \name to infer the correct ownership
scheme.

{\bf Translation.} With the above ownership scheme, \name performs the rewrites as in \autoref{fig:cell-crown} . Note that we do not attempt to rewrite \lstinline{name}
since it is an array pointer (see \autoref{sec:discussion} for limitations).

\section{Architecture} \label{sec:architecture}
In this section, we give a brief overview of \name's architecture.
\name takes as input Rust programs automatically translated by \crust.
These programs are very similar to the original C ones,
where the C syntax is replaced by Rust syntax.
\name applies several static analyses on the MIR of Rust
to infer properties of pointers:
\begin{itemize}

\item{\bf Ownership analysis}: computes ownership information about the pointers in the code, i.e. for each pointer it infers whether it is owning/non-owning at particular program locations.

\item{\bf Mutability analysis}: infers which pointers are used to modify the object they point to (inspired by \cite{10.1145/2384616.2384680,10.1145/1297027.1297051}).

\item{\bf Fatness analysis}: distinguishes array pointers from non-array pointers (inspired by \cite{10.1145/3527322}).
\end{itemize}  

The results of these analyses are summarised as type qualifiers~\cite{10.1145/1186632.1186635}.
A type qualifier is an atomic property (i.e., ownership, mutability, and fatness) that `qualifies' the standard pointer type.
These qualifiers are then utilised for pointer retyping. For example, an owning,
non-array pointer is retyped to \boxptr. %
After pointers have been retyped, \name rewrites their usages accordingly.

\section{Ownership Analysis} \label{sec:ownership}

The goal of our ownership analysis %
is to compute an ownership scheme for a given program that obeys the Rust ownership model, if such a scheme exists.
The ownership scheme contains information about whether pointers in the program are owning or non-owning at particular program locations.
At a high-level, our analysis works by generating a set of ownership constraints (section~\ref{sec:constraint-generation}),
which are then solved by a SAT solver (section~\ref{sec:constraint-solving}). A satisfying assignment for the
ownership constraints is an ownership scheme that obeys the Rust semantics.

Our ownership analysis is flow and field sensitive, where the latter enables inferring ownership information for pointer struct fields.
To satisfy field sensitivity, we track ownership information for \emph{access paths}~\cite{10.1007/978-3-642-31057-7_29,
10.1145/349299.349311,7372049}. 
Access paths represent a memory location by the way it is accessed from an initial, base variable,
and comprise of the base variable and a sequence of
field selection operators. %
For the program~\autoref{fig:llist-constr-rust}, some example access paths are \lstinline{new_node} (consists only of the base variable),
\lstinline{(*new_node).next}, and \lstinline{(*list).head}.
Our analysis associates an ownership variable with each access path, e.g. \lstinline{p} has associated ownership variable $\mathbb{O}_{p}$,
and  \lstinline{(*p).next} has associated ownership variable $\mathbb{O}_{(*p).next}$.
Each ownership variable can take value 1 if the corresponding access path is owning, or 0 if it is non-owning.
By ownership of an access path we mean the ownership of the field (or, more generally, pointer) accessed last through the access path,
e.g. the ownership of \lstinline{(*new_node).next} refers to the ownership of field \lstinline{next}.

\subsection{Ownership and aliasing} \label{sec:ownership-challenges}
One of the main challenges of designing an ownership analysis is the interaction between ownership and aliasing. To understand the problem, let us consider the pointer assignment at line 3 in the code listing below. We assume that the lines before the assignment allow inferring that \lstinline{q}, \lstinline{(*q).next} and \lstinline{r} are owning, whereas \lstinline{p} and \lstinline{(*p).next} are non-owning. Additionally, we assume that the lines after the assignment require \lstinline{(*p).next} to be owning (e.g. \lstinline{(*p).next} is being explicitly freed). From this, an ownership analysis
could reasonably conclude that ownership transfer happens at line 3 (such that \lstinline{(*p).next} becomes owning), and the inferred ownership scheme obeys the Rust semantics.
\begin{lstlisting}[language=rust, mathescape=true]
  let p, r, q : *mut Node;
  // p and (*p).next non-owning; q, (*q).next and r owning
  (*p).next = r; 
  // (*p).next must have ownership
\end{lstlisting}
Let's now also consider aliasing. A possible assumption is that, just before line 3,  \lstinline{p} and \lstinline{q} alias, meaning that \lstinline{(*p).next} and \lstinline{(*q).next} also alias.
Then, after line 3, \lstinline{(*p).next} and \lstinline{(*q).next} will still alias (pointing to the same memory object). However, according to the ownership scheme
above, both \lstinline{(*p).next} and \lstinline{(*q).next} are owning, which is not allowed in Rust, where a memory object must have a unique owner. This discrepancy was not detected by the ownership analysis mimicked above.
The issue is that the ownership analysis ignored aliasing. Indeed, ownership should not be transferred to \lstinline{(*p).next} if there exists an owning alias that, after the ownership transfer,
continues to point to the same memory object as \lstinline{(*p).next}.

Precise aliasing information is very difficult to compute, especially in the presence of inductively defined data structures.
In the current paper, we alleviate the need to check aliasing by making a strengthening assumption about the Rust ownership model: we restrict the way in which pointers can acquire ownership along an access path, thus limiting the interaction between ownership and aliasing.
In particular, we introduce a novel concept of \emph{ownership monotonicity}.
This property states that, along an access path, the ownership values of pointers can only decrease (see definition~\ref{def:monotonicity}, where $is\_prefix(p, q)$ returns true if access path $p$ is a prefix of $q$, and false otherwise -- e.g. $is\_prefix(p, (*p).next) = true$).
Going back to our example, the ownership monotonicity implies that, for access path \lstinline{(*p).next} we have $\mathbb{O}_{p} \ge \mathbb{O}_{(*p).next}$,
and for access path \lstinline{(*q).next} we have $\mathbb{O}_{q} \ge \mathbb{O}_{(*q).next}$.
This means that, if \lstinline{(*p).next} is allowed to take ownership, then \lstinline{p} must already be owning. Consequently, all aliases
of \lstinline{p} must be non-owning, which means that all aliases of \lstinline{(*p).next}, including \lstinline{(*q).next}, are non-owning.
\begin{definition}[Ownership monotonicity] \label{def:monotonicity}
  Given two access paths $p$ and $q$, if $is\_prefix(p, q)$, then $\mathbb{O}_{p} \geq \mathbb{O}_q$.
\end{definition}  
Ownership monotonicity is stricter than the Rust semantics, causing our analysis to potentially reject some ownership schemes that would otherwise be accepted by the
Rust compiler. We will come back to this point later in the section.

\subsection{Generation of ownership constraints} \label{sec:constraint-generation}

During constraint generation, we assume a given $k$ denoting the length of the longest access path used in the code.
This enables us to capture the ownership of all the access paths exposed in the code. Later in this section, we will discuss the handling of loops, which may expose longer access paths.

Next, we denote by $\mathcal{P}$ the set of all access paths in a program, $base\_var(p)$ returns the base variable of access path $p$,
and $|p|$ computes the length of the access path $p$ in terms of applied field selection operators from the base variable.
For illustration, $base\_var((*p).next) = p$, $base\_var(p) = p$, %
$|p| = 1$ and $|(*p).next| = 2$.
Then, we define $ap(v,lb, ub)$ to return the set of access paths with base variable $v$ and length in between lower bound $lb$ and upper bound $ub$:
  $ap(v, lb, ub) =  \left\{p \in \mathcal{P} \middle| base\_var(p) = v\land lb \le |p| \le ub\right\}$.
For illustration, we have 
$ap(new\_node, 1, 2) =  \left\{new\_node, (*new\_node).next\right\}$.

\vspace{-.5cm}
\begin{figure}
\begin{mathpar}
$
\inferrule[ASSIGN]
  {a = base\_var(v), ~~b=base\_var(w),\\
  p \in ap(a, |v|, k),~~ q \in ap(b,|w|, k), ~~r\in ap(a, 1, |v|{-}1), ~~s \in ap(b, 1, |w|{-}1)\\
  |p|-|v|=|q|-|w|, |r|=|s|\\
  C' = C \cup \{\mathbb{O}_{p}=0 \land \mathbb{O}_{p'}+\mathbb{O}_{q'}=\mathbb{O}_{q~} \land  \mathbb{O}_{r'}=\mathbb{O}_{r} \land \mathbb{O}_{s'}=\mathbb{O}_{s} \}}
  {C\vdash {\bf v = w;}\Rightarrow C'} 
$
\end{mathpar}
\vspace{-.5cm}
\caption{Ownership constraint generation for assignment}
\label{fig:ownership-rules}
\end{figure}
\vspace{-.5cm}

{\bf Ownership transfer.}
The program instructions where ownership transfer can happen are (pointer) assignment and function call. Here we discuss assignment and, due to space constraints, we leave the rules for interprocedural ownership analysis in Appendix~\ref{appendix:ownership}.
Our rule for ownership transfer at assignment site follows Rust's \boxptr semantics: when a \boxptr pointer is moved, the object it points to is moved as well.
For instance, in the following Rust pseudocode snippet:
\begin{lstlisting}
let p,q: Box<Box<i32>>;
let p = q; // ownership transfer occurs
// the use of q and *q is disallowed
\end{lstlisting}
when ownership is transferred from \lstinline{q} to \lstinline{p}, \lstinline{*q} also loses ownership.
Except for reassignment, the use of a \lstinline{Box} pointer after it lost its ownership is disallowed, hence the use of \lstinline{q} or \lstinline{*q} is forbidden at line 3.

Consequently, we enforce the following \emph{ownership transfer rule}: if ownership transfer happens for a pointer variable (e.g. \lstinline{p} and \lstinline{q} in the example), then it must happen for all pointers reachable   from that pointer (e.g. \lstinline{*p} and \lstinline{*q}). %
The ownership of pointer variables from which the pointer under discussion is reachable remains the same (e.g. if ownership transfer happens for some assignment \lstinline{*p = *q} in the code, then \lstinline{q} and
\lstinline{p} retain their respective previous ownership values).

\emph{Possible ownership transfer at pointer assignment:} The ownership transfer rule at pointer assignment site is captured by rule ASSIGN in~\autoref{fig:ownership-rules}.
The judgement $C\vdash {\bf v = w;}\Rightarrow C'$ denotes the fact that the assignment is analysed under the set of constraints $C$, and generates $C'$.
We use prime notation to denote variables after the assignment.
Given pointer assignment $v=w$, $p$ and $q$ represent all the access paths respectively starting from $v$ and $w$, whereas
$r$ and $s$ denote the access paths from the base variables of $v$ and $w$ that reach $v$ and $w$, respectively.
Then, equality $\mathbb{O}_{p'} + \mathbb{O}_{q'}=\mathbb{O}_{q}$ captures the possibility of ownership transfer for all access paths originating at $v$ and $w$:
(i) If transfer happens then the ownership of $q$ transfers to $p'$ ($\mathbb{O}_{p'}=\mathbb{O}_{q}$ and $\mathbb{O}_{q'}=0$).
(ii) Otherwise, the ownership values are left unchanged ($\mathbb{O}_{p'}=\mathbb{O}_{p}$ and $\mathbb{O}_{q'}=\mathbb{O}_{q}$).
The last two equalities, $ \mathbb{O}_{r'}=\mathbb{O}_{r} \land \mathbb{O}_{s'}=\mathbb{O}_{s}$, denote the fact that, for both (i) and (ii),
pointers on access paths $r$ and $s$ retain their previous ownership.

\emph{C memory leaks:} In the ASSIGN rule, we add constraint $\mathbb{O}_{p}=0$ to $C'$ in order to force $p$ to be non-owning before the assignment. Conversely, having $p$ owning before being reassigned via the assignment under analysis
signals a memory leak in the original C program. Given that in Rust memory is automatically returned, allowing the translation to happen would change the semantics of the original  
program by fixing the memory leak. Instead, our design choice is to disallow the ownership analysis from generating such a solution.

\emph{Simultaneous ownership transfer along an access path:} One may observe that the constraints generated by ASSIGN do not fully capture the stated ownership transfer rule. In particular,
they do not ensure that, whenever ownership transfer occurs from $w$ to $v$, it also transfers for all pointers on all access paths $p$ and $q$. Instead, this is implicitly guaranteed by the
ownership monotonicity rule, as stated in theorem~\ref{thm:transfer}.

\begin{theorem}[Ownership transfer]
  \label{thm:transfer}
  If ownership is transferred from $w$ to $v$, then, by the ASSIGN rule and ownership monotonicity,
  ownership also transfers between corresponding pointers on all access paths $p$ and $q$: $\mathbb{O}_{p'} =\mathbb{O}_{q}$ and $\mathbb{O}_{q'}=0$.  (proof in Appendix~\ref{sec:proof-of-transfer})
\end{theorem}

\emph{Ownership and aliasing:} We saw in section~\ref{sec:ownership-challenges} that aliasing may cause situations in which, after ownership transfer, the same memory object
has more than one owner. Theorem~\ref{thm:monotonicity} states that this is not possible under ownership monotonicity.

\begin{theorem}[Soundness of ASSIGN under ownership monotonicity] \label{thm:monotonicity}
  Under ownership monotonicity, if all allocated memory objects have a unique owner before the application of ASSIGN, then they will also have a unique owner after ASSIGN. (proof in Appendix~\ref{sec:proof-of-monotonicity})
\end{theorem}

Intuitively, theorem~\ref{thm:monotonicity} enables a pointer to acquire ownership without having to consider aliases: after ownership transfer, this pointer will be the unique owner.
The idea resembles that of strong updatess~\cite{10.1145/1926385.1926389}.

\emph{Additional access paths:} As a remark, it is possible for $v$ and $w$ to be accessible from other base variables in the program. 
In such cases, given that those access paths are not explicitly mentioned at the location of the ownership transfer, we do not generate new ownership variables for them.
Consequently, their current ownership variables are left unchanged by default.

{\bf Ownership transfer example.} To illustrate the ASSIGN rule, we use the singly-linked list example below, where we assume that \lstinline{p}, \lstinline{q} are both of
type \lstinline{*mut Node}. Therefore, we will have to consider the following four access path
\lstinline{p, q, (*p).next, (*q).next}.
In SSA-style, at each line in the example, we generate new ownership variables (by incrementing their subscript) for the access paths mentioned at that line.
For the first assignment, ownership transfer can happen between $p$ and $q$, and $(*p).next$ and $(*q).next$, respectively.
For the second assignment, ownership can be transferred between $(*p).next$ and $(*q).next$, while
$p$ and $q$ must retain their previous ownership.

\begin{lstlisting}[language=rust, mathescape=true]
p = q;    // $\mathbb{O}_{p_1} = 0 \land \mathbb{O}_{p_2} + \mathbb{O}_{q_2} = \mathbb{O}_{q_1}~\land$
          // $\mathbb{O}_{(*p_1).next} = 0 \land \mathbb{O}_{(*p_2).next} + \mathbb{O}_{(*q_2).next} = \mathbb{O}_{(*q_1).next}$
(*p).next = (*q).next;
          // $\mathbb{O}_{p_3} = \mathbb{O}_{p_2} \land \mathbb{O}_{q_3} = \mathbb{O}_{q_2}~\land$
          // $\mathbb{O}_{(*p_2).next} = 0 \land \mathbb{O}_{(*p_3).next} + \mathbb{O}_{(*q_3).next} = \mathbb{O}_{(*q_2).next}$

          
\end{lstlisting}

Besides generating ownership constraints for assignments, we must model
the ownership information for commonly used C standard function like \lstinline{malloc},
\lstinline{calloc}, \lstinline{realloc}, \lstinline{free}, \lstinline{strcmp}, \lstinline{memset}, etc..
Due to space constraints, more details about these, as well as the rules for ownership monotonicity and interprocedural ownership analysis are provided in Appendix~\ref{appendix:ownership}.

{\bf Handling conditionals and loops.}
As mentioned in section~\ref{sec:overview-loop}, we only analyse the body of loops once as it is sufficient to expose  all the required ownership variables.
For inductively defined data structures, while further unrolling of loop bodies increases the length of access paths, it does not expose
any new struct fields (struct fields do not change ownership between loop iterations).

To handle join points of control paths, we apply a variant of the SSA  construction algorithm~\cite{10.5555/295545.295551},
where different paths are merged via $\phi$ nodes. The value of each ownership variable must be the same
on all joined paths, or otherwise the analysis fails. %

\subsection{Solving ownership constraints} \label{sec:constraint-solving}

The ownership constraint system consists of a set of 3-variable linear constraints of the form $O_v = O_w + O_u$, and
1-variable equality constraints $O_v=0$ and $O_v=1$. %

\begin{definition}[Ownership constraint system]
An \emph{ownership constraint system} $(P, \Delta, \Sigma, \Sigma_{\neg})$ consists of a set of ownership 
variables $P$ that can have either value 0 or 1, %
a set of 3-variable equality constraints $\Delta \subseteq P \times P \times P$, and two sets of 1-variable equality 
constraints, $\Sigma, \Sigma_{\neg} \subseteq P$.
The equalities in $\Sigma$ are of the form $x=1$, whereas the equalities in $\Sigma_{\neg}$ are of the form $x=0$.%
\end{definition}

\begin{theorem}[Complexity of the ownership constraint solving]\label{thm:ownership}
The ownership constraint solving is NP-complete.  (proof in Appendix~\ref{appendix:thm-proof})
\end{theorem}

We solve the ownership constraints by calling a SAT solver. 
The ownership constraints may have no solution. This happens when there is no ownership scheme that obeys the Rust ownership model and the ownership monotonicity property (which is stricter than the Rust model for some cases), or the original C program has a memory leak.

Due to the complex Rust semantics, we do not formally prove that a satisfying assignment obeys the Rust ownership model. Instead, this check is performed
after the translation by running the Rust compiler.

\subsection{Discussion on ownership monotonicity} \label{sec:ownership-monotonicity}

As mentioned earlier in section~\ref{sec:ownership-monotonicity}, ownership monotonicity is stricter than the Rust semantics, causing our analysis to potentially reject some ownership schemes that would otherwise be accepted by the Rust compiler. The most significant such scenario that we identified is that of \emph{reference output parameter}. This denotes a function
parameter passed as a reference, which acts as an output as it can be accessed from outside the function (e.g. \lstinline{list} in figure~\ref{fig:llist-constr-c}).
For such parameters, the base variable is non-owning (as it is a reference) and mutable, whereas 
the pointers reachable from it may be owning (see example in figure~\ref{fig:llist-constr-crown}, where \lstinline{(*node).head}
gets assigned a pointer to a  newly allocated node). 
We detect such situations and explicitly enable them. In particular, we explicitly convert owning pointers \lstinline{p} to \lstinline{&mut(*p)} at the translation stage.

\section{C to Rust Translation} \label{sec:translation}
\name uses the results of the ownership, mutability and fatness analyses
to perform the actual translation, which consists of retyping pointers (section~\ref{sec:pointer-retype})
and rewriting pointer uses (section~\ref{sec:pointer-rewrite}).

\subsection{Retyping pointers} \label{sec:pointer-retype}

As mentioned in section~\ref{sec:backgroud-pointers}, we do not attempt to translate array pointers to safe pointers.
In the rest of the section, we focus on mutable, non-array pointers.

The translation requires a global view of pointers' ownership, whereas
information inferred by the ownership analysis refers to individual program locations.
For the purpose of translation, given that we refactor owning pointers into box pointers,
a pointer is considered (globally) owning if it owns a memory object at any program location within its scope.
Otherwise, it is (globally) non-owning.
When retyping pointer fields of structs, we must consider the scope of the struct declaration, which
generally transcends the whole program. Within this scope, each field is usually accessed
from several base variables, which must all be taken into consideration. For instance, given the
\lstinline{List} declaration in figure~\ref{fig:llist-constr-rust} and two variables \lstinline{l1}
and \lstinline{l2} of type \lstinline{*mut List}.
Then, in order to determine the ownership status of field \lstinline{next}, we have to consider all the access paths to \lstinline{next} originating from both base variables
\lstinline{l1} and \lstinline{l2}.

The next table shows the retyping rules for mutable, non-array pointers, where we wrap safe pointer types into \lstinline{Option} to account for null pointer values:
\begin{center}
    \begin{tabular}{ |c|c| } 
    \hline
     & Non-array pointers \\
    \hline
    Owning & \optionboxptr{T} \\
    \hline
    Non-owning &  \rawmut{T} or \optionmutref{T} \\ %
    \hline
    \end{tabular}
\end{center}

The non-owning pointers that are kept as raw pointers \rawmut{T} correspond to 
mutable local borrows. For \name to soundly translate to mutable local borrows, it would have to reason about lifetime information.
Given that in Rust there can't be two mutable references to a value, \name would have to check
that there is no overlap between the lifetimes of different mutable references to the same object. 
In this work, we chose not to do this and instead leave it as a future work (as also mentioned under limitations in section~\ref{sec:discussion}).
Notably, this restriction does not apply to output parameters, for which we translate to mutable references.
This already covers the majority of mutable references -- Das observed that, in C code, most references arise because the address of a variable is passed as a
parameter~\cite{10.1145/349299.349309}. Notably, the lack of a lifetime analysis means that we also can't handle immutable local borrows, hence our translation's focus on
mutable pointers.

\subsection{Rewriting pointer uses}\label{sec:pointer-rewrite}

The rewrite of a pointer expression depends on its new type and the context in which it is used.
For example, when rewriting \lstinline{q} in \lstinline{p = q}, the context will depend on the new
type of \lstinline{p}. Based on this new type, we can have four contexts: $\textsf{BoxCtxt}$ which requires \boxptr pointers, 
	$\textsf{MutCtxt}$ which requires \mutref references, $\textsf{ConstCtxt}$ which requires \constref references, and
    $\textsf{RawCtxt}$ which requires raw pointers.
For example, if \lstinline{p} above is a \boxptr pointer, then we rewrite \lstinline{q} in a $\textsf{BoxCtxt}$.

Then, the rewrite takes place according to the following table, where columns correspond to the new type of the pointer to be rewritten, and rows represent possible contexts
\footnote{The cell marked as $\bot$ is not applicable due to our treatment of output parameter.}.

\begin{centering}
  \begin{tabular}{ |c|c|c|c| } 
    \hline
     & \optionboxptr{T} & \optionmutref{T} & \rawmut{T} \\
    \hline
    \textsf{BoxCtxt} & \lstinline[]$p.take()$ & $\bot$ & \lstinline[]$Some(Box::from_raw(p))$ \\
    \hline
    \textsf{MutCtxt} & \lstinline[]$p.as_deref_mut()$ & \lstinline[]$p.as_deref_mut()$ &  \lstinline[]$p.as_mut()$ \\
    \hline
    \textsf{ConstCtxt} & \lstinline[]$p.as_deref()$ & \lstinline[]$p.as_deref()$ &  \lstinline[]$p.as_ref()$ \\
    \hline
    \textsf{RawCtxt} & \lstinline[]$to_raw(&mut p)$ & \lstinline[]$to_raw(&mut p)$ &  \lstinline[]$p$ \\
    \hline
  \end{tabular}
\end{centering}

Our translation uses functions from the Rust standard library, as follows:
\begin{enumerate}
    \item When \lstinline{Option<Box<T>>} is passed to a $\textsf{BoxCtxt}$, we expect a move, and consequently we use 
    \lstinline{take} to replace the value inside the option with None;
    \item We use \lstinline{as_deref} and \lstinline{as_deref_mut} in order to not 
    consume the original option, and we create new options with references to the original ones;
    \item \lstinline{as_mut} and \lstinline{as_ref} converts raw pointers to references;
    \item \lstinline{Box::from_raw} converts raw pointers into \boxptr pointers.
\end{enumerate}

We also define the helper function \lstinline{to_raw} that transform safe pointers into raw pointers:
\begin{lstlisting}[basicstyle=\ttfamily\small,numbers=none]
    fn to_raw<T>(b: &mut Option<Box<T>>) -> *mut T {
        b.as_deref_mut().map(|b| b as *mut T).unwrap_or(null_mut())
    }
\end{lstlisting}
Here, we explain \lstinline{to_raw} for a \boxptr argument (the explanation for \mutref is the
same because of the polymorphic nature of \lstinline{as_deref_mut}): 
\begin{enumerate}
    \item To convert \lstinline{Option<Box<T>>}, we first mutably borrow the entire option as denoted 
    by the mutable borrow argument of the helper function. This is needed because \lstinline{Option} is
    not copyable, and it would be otherwise consumed;
    \item \lstinline{as_deref_mut} converts \lstinline{&mut Option<Box<T>>} to \lstinline{Option<&mut T>};
    \item \lstinline{map} converts the optional part of the reference
    into an option of raw pointers;
    \item Finally, \lstinline{unwrap_or} returns the \lstinline{Some} value of
    the option, or a null pointer \lstinline{std::ptr::null_mut()} if the value is \lstinline{None}.
\end{enumerate}

\emph{Dereferences:} When a pointer \lstinline{p} is dereferenced as part of a larger expression (e.g. \lstinline{(*p).next}),
we need an additional \lstinline{unwrap()}.

\emph{Box pointers check:} Rust disallows the use of \lstinline{Box} pointers after they lost their ownership.
As this rule cannot be captured by the ownership analysis, such situations are detected at translation stage,
and the culpable \lstinline{Box} pointers are reverted back to raw pointers.

For brevity, we omitted the slightly different treatment of struct fields that are not of pointer type.

\section{Challenges of Handling Real-World Code} \label{sec:discussion}
We designed \name to be able to analyse and translate real-world code,
which poses significant challenges. In this section, we discuss some of the engineering challenges of \name and its current limitations. 

\subsection{Preprocessing}
During the transpilation of  C libraries,  \crust treats each file as a separate compilation
unit, which gets translated into a separate Rust module. Consequently, struct
definitions are duplicated, and available function definitions are put in \lstinline{extern} blocks~\cite{DBLP:journals/pacmpl/EmreSDH21}. 
We apply a preprocessing step similar to the resolve-imports tool of Laertes~\cite{DBLP:journals/pacmpl/EmreSDH21} that links
those definitions across files.

\subsection{Limitations of the ownership analysis}
There are a few C constructs and idioms that are not fully supported by our implementation, for which \name generates partial ownership constraints.
\name's translation will attempt to rewrite a variable as long as there exists a constraint involving it. 
As a result, the translation is in theory neither \emph{sound} nor \emph{complete}: it may generate code that does not compile (though we have not observed this in practice~(see Section \ref{sec:experiments})) and it may leave some pointers as raw pointers resulting in a less than optimal translation. 
We list below the cases when such a scenario may happen. \vspace{-0.2cm}

\paragraph{Certain unsafe C constructs.} %
  For type casts, we only generate ownership transfer constraints for head pointers;
for unions we assume that they contain no pointer fields and consequently, we generate no constraints; similarly, we generate no constraints for variadic arguments.
We noticed that unions and variadic arguments may 
cause our tool to crash (e.g. three of the benchmarks in \cite{DBLP:journals/pacmpl/EmreSDH21},
as mentioned in Section \ref{sec:experiments}). Those crashes happen when analysing access paths
that contain dereferences of union fields (where we assumed no pointer fields), and when analysing calls to functions with variadic arguments where a pointer is passed as argument.\vspace{-0.2cm}
\paragraph{Function pointers.} \name does not generate any constraints for them. \vspace{-0.2cm}

\paragraph{Non-standard memory management in C libraries.} Certain C libraries %
wrap \lstinline{malloc} and \lstinline{free}, often with static function pointers (pointers to allocator/deallocator are stored in static variables),
or function pointers in structs. \name does not generate any constraints in such scenarios. %
In C, it is also possible to use \lstinline{malloc} to allocate a large
piece of memory, and then split it into several sub-regions assigned to different pointers.
In our ownership analysis, only one pointer can gain ownership of the memory allocated by a call to \lstinline{malloc}.
Another C idiom that we don't fully support %
occurs when certain pointers can 
point to either heap allocated objects, or statically allocated stack arrays.
\name generates ownership constraints only for the heap and, consequently, those variables will be left under-constrained. %

\subsection{Other limitations of \name}

\paragraph{Array pointers.} For array pointers, although \name infers the correct ownership information, it does not generate the meta data required to synthesise Rust code.\vspace{-0.2cm}

\paragraph {Mutable local borrows.} As explained in the last paragraph of Section~\ref{sec:pointer-retype},
\name does not translate mutable non-owning pointers to local mutable references as this requires dedicated analysis of lifetimes. 
Note that \name does however generate mutable references for output parameters.\vspace{-0.2cm}

\paragraph{Access paths that break ownership monotonicity.}
As discussed in section~\ref{sec:ownership-monotonicity}, ownership monotonicity may be stricter in certain cases than Rust's semantics.

\section{Experimental Evaluation} \label{sec:experiments}
We implement \name on top of the Rust compiler, version \texttt{nightly-2023-01-26}. We use
\crust with version 0.16.0. For the SAT solver, we rely on a Rust-binding of \texttt{z3}\cite{z3}
with version 0.11.2.
We run all our experiments on a MacBook Pro with an Apple M1 chip, with 8 cores (4 performance
and 4 efficiency). The computer has 16GB RAM and runs macOS Monterey 12.5.1.

{\bf Benchmark selection.}
To evaluate the utility of \name, we collected a benchmark suite of 20 programs (Table~\ref{tab:benchmarks}).
These include benchmarks from Laertes~\cite{DBLP:journals/pacmpl/EmreSDH21}'s accompanying 
artifact~\cite{emre_mehmet_2021_5442253} (marked by * in Table~\ref{tab:benchmarks})\footnote{We excluded \texttt{json-c}, \texttt{optipng}, \texttt{tinycc}
where \name crashes because
of the uses of unions and variadic arguments as discussed in Section~\ref{sec:discussion}. Additional programs (\texttt{qsort}, \texttt{grabc}, \texttt{xzoom}, \texttt{snudown}, \texttt{tmux},
\texttt{libxml2}) are mentioned in the paper~\cite{DBLP:journals/pacmpl/EmreSDH21} but are either missing 
or incomplete in the artifact~\cite{emre_mehmet_2021_5442253}.}, and additionally 8 real-world projects (\texttt{binn}, \texttt{brotli}, \texttt{buffer},
\texttt{heman}, \texttt{json.h}, \texttt{libtree}, \texttt{lodepng}, \texttt{rgba}) together with 4 commonly used data structure libraries (\texttt{avl}, \texttt{bst}, \texttt{ht}, \texttt{quadtree}).

\vspace{-1cm}
\begin{table}[!ht]
\caption{Benchmark information}
    \centering
    \begin{tabular}{lrrrr|lrrrr}
        Benchmark & Files         & Structs          & Functions            & LOC    & Benchmark         & Files         & Structs          & Functions            & LOC   \\
        avl       & 1             & 2                & 11                   & 229    & libcsv*           & 1             & 6                & 23                   & 976   \\
        binn      & 1             & 5                & 165                  & 4426   & libtree           & 1             & 18               & 32                   & 2610  \\
        brotli    & 30            & 237              & 867                  & 537723 & libzahl*          & 49            & 65               & 108                  & 4655  \\
        bst       & 1             & 1                & 6                    & 154    & lil*              & 2             & 9                & 136                  & 5670  \\
        buffer    & 2             & 3                & 42                   & 1207   & lodepng           & 1             & 19               & 236                  & 14153 \\
        bzip2*    & 9             & 39               & 126                  & 14829  & quadtree          & 5             & 14               & 31                   & 1216  \\
        genann*   & 6             & 10               & 27                   & 2410   & rgba              & 2             & 3                & 19                   & 1855  \\
        heman     & 24            & 52               & 302                  & 13762  & robotfindskitten* & 1             & 8                & 18                   & 1508  \\
        ht        & 1             & 3                & 10                   & 264    & tulipindicators*  & 111           & 18               & 229                  & 22363 \\
        json.h    & 1             & 13               & 53                   & 3860   & urlparser*        & 1             & 1                & 21                   & 1379 
        \end{tabular}
    \label{tab:benchmarks}
\end{table}
\vspace{-1cm}

\subsection{Research questions}
We aim at answering the following research questions.
\begin{itemize}
    \item[] RQ1. How many raw pointers/pointer uses can \name translate to safe pointers/pointer uses?
    \item[] RQ2. How does \name's result compare with the state-of-the-art~\cite{DBLP:journals/pacmpl/EmreSDH21}?
    \item[] RQ3. What is the runtime performance of \name?
\end{itemize}

{\bf RQ 1: Unsafe pointer reduction.}
In order to judge \name's efficacy, we measure the reduction rate of raw pointer declarations and uses.
This is a direct indicative of the improvement in safety, as safe pointers are always checked by the Rust compiler (even inside \lstinline{unsafe} regions).  
As previously mentioned, we focus on mutable non-array pointers. %
The results are presented in Table~\ref{tab:unsafe-reduce},
where \#ptrs counts the number of raw pointer declarations in a given benchmark, \#uses counts the
number of times raw pointers are being used, and the Laertes and Crown headers denote the reduction rates of the
number of raw pointers and raw pointer uses achieved by the two tools, respectively.
For instance, for benchmark \texttt{avl}, the rate of 100\% means that 
all raw pointer declarations and their uses are translated into safe ones.
Note that the ``-'' symbols on the row corresponding to \lstinline{robotfindskitten}
are due to the fact that the benchmark contains 0 raw pointer uses.

The median reduction rates achieved by \name for raw pointers and raw pointer uses are
37.3\% and 62.1\%, respectively. 
\name achieves a 100\% reduction rate for many non-trivial data
structures (\lstinline{avl}, \lstinline{bst}, \lstinline{buffer}, \lstinline{ht}), as well as for \lstinline{rgba}. 
For \texttt{brotli}, a lossless data compression algorithm developed by Google, which is our largest benchmark,  
\name achieves reduction rates of 21.4\% and 20.9\%, respectively.  
The relatively low reduction rates for \lstinline{brotli} and a few other  
benchmarks (\texttt{tulipindicators}, \texttt{lodepng}, \texttt{bzip2}, \texttt{genann},
\texttt{libzahl}) is due to their use of non-standard memory management strategies (discussed in detail in
Section \ref{sec:discussion}). %

Notably, all the translated benchmarks compile under the aforementioned Rust compiler version.
As a check of semantics preservation, for the benchmarks that provide test suites
(\lstinline{libtree}, \lstinline{rgba}, \lstinline{quadtree}, \lstinline{urlparser}, \lstinline{genann}, \lstinline{buffer}),
our translated benchmarks pass all the provided tests.

\vspace{-.9cm}
\begin{table}[]
    \centering
\caption{Reduction of (mutable, non-array) raw pointer declarations and uses}
\begin{tabular*}{\textwidth}{c @{\extracolsep{\fill}} lrrrrrr}
    Benchmark         & \#ptrs & Laertes         & Crown            & \#uses & Laertes & Crown            \\
    avl               & 8      & 0.0\%           & \textbf{100.0\%} & 41     & 0.0\%   & \textbf{100.0\%} \\
    binn              & 103    & 46.6\%          & \textbf{65.0\%}  & 247    & 62.3\%  & \textbf{71.3\%}  \\
    brotli            & 846    & 0.0\%           & \textbf{21.4\%}  & 3686   & 0.0\%   & \textbf{20.9\%}  \\
    bst               & 5      & 0.0\%           & \textbf{100.0\%} & 22     & 0.0\%   & \textbf{100.0\%} \\
    buffer            & 38     & 0.0\%           & \textbf{100.0\%} & 56     & 0.0\%   & \textbf{100.0\%} \\
    bzip2*            & 126    & 14.3\%          & \textbf{26.2\%}  & 2946   & 2.2\%   & \textbf{3.7\%}   \\
    genann*           & 28     & 0.0\%           & \textbf{7.1\%}   & 160    & 0.0\%   & \textbf{15.0\%}  \\
    heman             & 360    & 30.3\%          & \textbf{35.0\%}  & 926    & 50.2\%  & \textbf{60.2\%}  \\
    ht                & 6      & 33.3\%          & \textbf{100.0\%} & 28     & 42.9\%  & \textbf{100.0\%} \\
    json.h            & 128    & 2.3\%           & \textbf{23.4\%}  & 647    & 1.2\%   & \textbf{62.1\%}  \\
    libcsv*           & 20     & 65.0\%          & \textbf{70.0\%}  & 141    & 97.9\%  & 97.9\%           \\
    libtree           & 48     & 29.2\%          & \textbf{39.6\%}  & 227    & 33.0\%  & \textbf{62.1\%}  \\
    libzahl*          & 87     & 2.2\%           & \textbf{16.1\%}  & 279    & 4.1\%   & \textbf{16.8\%}  \\
    lil*              & 202    & 9.2\%           & \textbf{18.8\%}  & 1018   & 51.4\%  & \textbf{69.4\%}  \\
    lodepng           & 227    & \textbf{46.3\%} & 44.9\%           & 1232   & 40.4\%  & 37.7\%           \\
    quadtree          & 33     & 0.0\%           & \textbf{42.4\%}  & 117    & 0.0\%   & \textbf{48.7\%}  \\
    rgba              & 6      & 83.3\% & 83.3\%  & 12     & 100.0\% & 100.0\%          \\
    robotfindskitten* & 1      & 0.0\%           & 0.0\%            & 0      & -       & -                \\
    tulipindicators*  & 134    & 0.0\%           & \textbf{0.7\%}   & 625    & 0.0\%   & 0.0\%            \\
    urlparser*        & 9      & 0.0\%           & \textbf{11.1\%}  & 40     & 0.0\%   & \textbf{45.0\%} 
\end{tabular*}
\label{tab:unsafe-reduce}
\end{table}
\vspace{-.7cm}

{\bf RQ 2: Comparing with state-of-the-art.}
The comparison of \name with Laertes~\cite{DBLP:journals/pacmpl/EmreSDH21}
is also shown in Table \ref{tab:unsafe-reduce}, with bold font highlighting
better results. The data on Laertes is either directly extracted from 
the artifact~\cite{emre_mehmet_2021_5442253} or has been confirmed by 
the authors through private correspondence. We can see that \name outperforms 
the state-of-the-art (often by a significant degree) in most cases, with \texttt{lodepng} being the only exception, where we suspect that the reason also
lies with non-standard memory management strategies mentioned before. Laertes
is less affected by this as it does not rely on ownership analysis.

{\bf RQ 3: Runtime performance.}
Although our analysis relies on solving a constraint satisfaction problem that is proven to
be NP-complete, in practice the runtime performance of \name is consistently  high.
The execution time of the analysis and the rewrite for the whole benchmark suite is within 60 seconds.

\section{Related Works}
{\bf Ownership analysis.} The concept of ownership itself is not new. 
In OO programming, ownership type systems are used to enable controlled aliasing by
restricting object graphs 
underlying the runtime heap of object-oriented programs~\cite{KULeuven-2-1630873, 10.1145/286936.286947}.
Efforts have been made in the automatic inference of ownership types~\cite{10.1145/582419.582448,10.1145/3485522,10.1145/604131.604156},
and applications of ownership types for memory management
~\cite{10.1145/781131.781168,ZHAO2008213}. Similarly, the concept of ownership has also been applied to analyse \texttt{C/C++} programs. 
Heine et al.~\cite{DBLP:conf/pldi/HeineL03} inferred pointer ownership type for memory leak
detection. Their encoding of ownership transfer constraints greatly influenced us.
Ravitch et al. \cite{pldi09} apply static analysis to infer ownership for automatic library
binding generation. Our notion of output parameter is inspired by this work.
Giving the different application domains, each of these work makes different
assumptions. 
Heine et al.~\cite{DBLP:conf/pldi/HeineL03} assumes that indirectly-accessed pointers cannot acquire ownership, whereas 
Ravitch et al. \cite{pldi09} assumes that all struct fields are owning unless
explicitly annotated. Our analysis is free from these assumptions, which 
leads to a more precise analysis necessary for synthesising correct Rust code.   
Lastly, the idea of ownership is also broadly applied in concurrent separation logic~\cite{10.1016/j.tcs.2006.12.034,10.1016/j.entcs.2006.04.008,
10.1109/LICS.2007.30,10.1145/1480881.1480922,10.1145/1122971.1122992}.
The Iris framework~\cite{jung:hal-01945446} is also applied to formalise the Rust type 
system~\cite{10.1145/3158154} and verifying Rust programs~\cite{10.1145/3519939.3523704}.

{\bf Type qualifiers.} Type qualifiers are a lightweight, practical mechanism for 
specifying and checking properties not captured by traditional type systems. A general
flow-insensitive type qualifier framework has been proposed \cite{10.1145/1186632.1186635},
with subsequent applications analysing \texttt{Java} reference 
mutability~\cite{10.1145/1297027.1297051,10.1145/2384616.2384680} and \texttt{C} array bounds~\cite{10.1145/3527322}. 
Our mutability and fatness analyses of \name are greatly influenced by these work.

{\bf C to Rust Translation.} 
We have already discussed \crust~\cite{c2rust}, which is an industrial strength tool that converts C to Rust syntax. \crust does not attempt to fix unsafe features such as raw pointers and the programs it generates are always annotated as unsafe. Nevertheless it forms the bases of other translation efforts. CRustS~\cite{CRustS} applies AST-based code transformations
to remove superfluous unsafe labelling generated by \crust. But it does not 
fix the unsafe features either. Laertes~\cite{DBLP:journals/pacmpl/EmreSDH21}
is the first tool that is actually able to automatically reduce the presence of unsafe code. 
It uses the Rust compiler as a blackbox oracle and search for code changes that 
remove raw pointers, which is very different from \name in approach (see Section~\ref{sec:experiments} for an experimental comparison).

\section{Conclusion} \label{sec:conclusion}
We devised an ownership analysis for Rust programs translated by \crust that 
is scalable (handling half a million LOC in less than 10 seconds) 
and precise (handling inductive data structures) thanks to a strengthening of the Rust ownership model, which we call 
ownership monotonicity. 
Based on this new analysis, we prototyped a refactoring tool for translating C programs
into Rust programs. Our experimental evaluation shows that the proposed approach handles real-world benchmarks and
outperforms the state-of-the-art.

\bibliographystyle{splncs04}
\bibliography{paper.bib}

\begin{thebibliography}{10}
\providecommand{\url}[1]{\texttt{#1}}
\providecommand{\urlprefix}{URL }
\providecommand{\doi}[1]{https://doi.org/#1}

\bibitem{10.1145/582419.582448}
Aldrich, J., Kostadinov, V., Chambers, C.: Alias annotations for program
  understanding. In: Proceedings of the 17th ACM SIGPLAN Conference on
  Object-Oriented Programming, Systems, Languages, and Applications. p.
  311–330. OOPSLA '02, Association for Computing Machinery, New York, NY, USA
  (2002). \doi{10.1145/582419.582448},
  \url{https://doi.org/10.1145/582419.582448}

\bibitem{10.1145/604131.604156}
Boyapati, C., Liskov, B., Shrira, L.: Ownership types for object encapsulation.
  In: Proceedings of the 30th ACM SIGPLAN-SIGACT Symposium on Principles of
  Programming Languages. p. 213–223. POPL '03, Association for Computing
  Machinery, New York, NY, USA (2003). \doi{10.1145/604131.604156},
  \url{https://doi.org/10.1145/604131.604156}

\bibitem{10.1145/781131.781168}
Boyapati, C., Salcianu, A., Beebee, W., Rinard, M.: Ownership types for safe
  region-based memory management in real-time java. In: Proceedings of the ACM
  SIGPLAN 2003 Conference on Programming Language Design and Implementation. p.
  324–337. PLDI '03, Association for Computing Machinery, New York, NY, USA
  (2003). \doi{10.1145/781131.781168},
  \url{https://doi.org/10.1145/781131.781168}

\bibitem{10.5555/295545.295551}
Briggs, P., Cooper, K.D., Harvey, T.J., Simpson, L.T.: Practical improvements
  to the construction and destruction of static single assignment form. Softw.
  Pract. Exper.  \textbf{28}(8),  859–881 (jul 1998)

\bibitem{10.1016/j.entcs.2006.04.008}
Brookes, S.: Variables as resource for shared-memory programs: Semantics and
  soundness. Electron. Notes Theor. Comput. Sci.  \textbf{158},  123–150 (may
  2006). \doi{10.1016/j.entcs.2006.04.008},
  \url{https://doi.org/10.1016/j.entcs.2006.04.008}

\bibitem{10.1016/j.tcs.2006.12.034}
Brookes, S.: A semantics for concurrent separation logic. Theor. Comput. Sci.
  \textbf{375}(1–3),  227–270 (apr 2007). \doi{10.1016/j.tcs.2006.12.034},
  \url{https://doi.org/10.1016/j.tcs.2006.12.034}

\bibitem{10.1109/LICS.2007.30}
Calcagno, C., O'Hearn, P.W., Yang, H.: Local action and abstract separation
  logic. In: Proceedings of the 22nd Annual IEEE Symposium on Logic in Computer
  Science. p. 366–378. LICS '07, IEEE Computer Society, USA (2007).
  \doi{10.1109/LICS.2007.30}, \url{https://doi.org/10.1109/LICS.2007.30}

\bibitem{10.1145/349299.349311}
Cheng, B.C., Hwu, W.M.W.: Modular interprocedural pointer analysis using access
  paths: Design, implementation, and evaluation. In: Proceedings of the ACM
  SIGPLAN 2000 Conference on Programming Language Design and Implementation. p.
  57–69. PLDI '00, Association for Computing Machinery, New York, NY, USA
  (2000). \doi{10.1145/349299.349311},
  \url{https://doi.org/10.1145/349299.349311}

\bibitem{KULeuven-2-1630873}
Clarke, D., {\"O}stlund, J., Sergey, I., Wrigstad, T.: Ownership types: a
  survey. vol.~7850, pp. 15--58. Springer (2013).
  \doi{https://doi.org/10.1007/978-3-642-36946-9-3},
  \url{https://lirias.kuleuven.be/1630873}

\bibitem{10.1145/286936.286947}
Clarke, D.G., Potter, J.M., Noble, J.: Ownership types for flexible alias
  protection. In: Proceedings of the 13th ACM SIGPLAN Conference on
  Object-Oriented Programming, Systems, Languages, and Applications. p.
  48–64. OOPSLA '98, Association for Computing Machinery, New York, NY, USA
  (1998). \doi{10.1145/286936.286947},
  \url{https://doi.org/10.1145/286936.286947}

\bibitem{10.1145/349299.349309}
Das, M.: Unification-based pointer analysis with directional assignments. In:
  Proceedings of the ACM SIGPLAN 2000 Conference on Programming Language Design
  and Implementation. p. 35–46. PLDI '00, Association for Computing
  Machinery, New York, NY, USA (2000). \doi{10.1145/349299.349309},
  \url{https://doi.org/10.1145/349299.349309}

\bibitem{10.1007/978-3-642-31057-7_29}
De, A., D'Souza, D.: Scalable flow-sensitive pointer analysis for java with
  strong updates. In: Noble, J. (ed.) ECOOP 2012 -- Object-Oriented
  Programming. pp. 665--687. Springer Berlin Heidelberg, Berlin, Heidelberg
  (2012)

\bibitem{emre_mehmet_2021_5442253}
Emre, M., Schroeder, R.: Artifact for "translating c to safer rust" (Sep 2021).
  \doi{10.5281/zenodo.5442253}, \url{https://doi.org/10.5281/zenodo.5442253}

\bibitem{DBLP:journals/pacmpl/EmreSDH21}
Emre, M., Schroeder, R., Dewey, K., Hardekopf, B.: Translating {C} to safer
  rust. Proc. {ACM} Program. Lang.  \textbf{5}({OOPSLA}),  1--29 (2021).
  \doi{10.1145/3485498}, \url{https://doi.org/10.1145/3485498}

\bibitem{10.1145/1480881.1480922}
Feng, X.: Local rely-guarantee reasoning. In: Proceedings of the 36th Annual
  ACM SIGPLAN-SIGACT Symposium on Principles of Programming Languages. p.
  315–327. POPL '09, Association for Computing Machinery, New York, NY, USA
  (2009). \doi{10.1145/1480881.1480922},
  \url{https://doi.org/10.1145/1480881.1480922}

\bibitem{z3}
Fitzgerald, N., Hoare, G., Mitchener, B., Puri, S.: Rust bindings to the z3 smt
  solver. \url{https://crates.io/crates/z3}

\bibitem{10.1145/1186632.1186635}
Foster, J.S., Johnson, R., Kodumal, J., Aiken, A.: Flow-insensitive type
  qualifiers. ACM Trans. Program. Lang. Syst.  \textbf{28}(6),  1035–1087
  (nov 2006). \doi{10.1145/1186632.1186635},
  \url{https://doi.org/10.1145/1186632.1186635}

\bibitem{10.1145/1297027.1297051}
Greenfieldboyce, D., Foster, J.S.: Type qualifier inference for java. In:
  Proceedings of the 22nd Annual ACM SIGPLAN Conference on Object-Oriented
  Programming Systems, Languages and Applications. p. 321–336. OOPSLA '07,
  Association for Computing Machinery, New York, NY, USA (2007).
  \doi{10.1145/1297027.1297051}, \url{https://doi.org/10.1145/1297027.1297051}

\bibitem{DBLP:conf/pldi/HeineL03}
Heine, D.L., Lam, M.S.: A practical flow-sensitive and context-sensitive {C}
  and {C++} memory leak detector. In: Cytron, R., Gupta, R. (eds.) Proceedings
  of the {ACM} {SIGPLAN} 2003 Conference on Programming Language Design and
  Implementation 2003, San Diego, California, USA, June 9-11, 2003. pp.
  168--181. {ACM} (2003). \doi{10.1145/781131.781150},
  \url{https://doi.org/10.1145/781131.781150}

\bibitem{10.1145/2384616.2384680}
Huang, W., Milanova, A., Dietl, W., Ernst, M.D.: Reim \& reiminfer: Checking
  and inference of reference immutability and method purity. In: Proceedings of
  the ACM International Conference on Object Oriented Programming Systems
  Languages and Applications. p. 879–896. OOPSLA '12, Association for
  Computing Machinery, New York, NY, USA (2012). \doi{10.1145/2384616.2384680},
  \url{https://doi.org/10.1145/2384616.2384680}

\bibitem{c2rust}
inc., I.: c2rust. \url{https://github.com/immunant/c2rust}

\bibitem{10.1145/3158154}
Jung, R., Jourdan, J.H., Krebbers, R., Dreyer, D.: Rustbelt: Securing the
  foundations of the rust programming language. Proc. ACM Program. Lang.
  \textbf{2}(POPL) (dec 2017). \doi{10.1145/3158154},
  \url{https://doi.org/10.1145/3158154}

\bibitem{jung:hal-01945446}
Jung, R., Krebbers, R., Jourdan, J.H., Bizjak, A., Birkedal, L., Dreyer, D.:
  {Iris from the ground up: A modular foundation for higher-order concurrent
  separation logic}. {Journal of Functional Programming}  \textbf{28}(e20)
  (2018). \doi{10.1017/S0956796818000151},
  \url{https://hal.science/hal-01945446}

\bibitem{7372049}
Lerch, J., Späth, J., Bodden, E., Mezini, M.: Access-path abstraction: Scaling
  field-sensitive data-flow analysis with unbounded access paths (t). In: 2015
  30th IEEE/ACM International Conference on Automated Software Engineering
  (ASE). pp. 619--629 (2015). \doi{10.1109/ASE.2015.9}

\bibitem{10.1145/1926385.1926389}
Lhot\'{a}k, O., Chung, K.C.A.: Points-to analysis with efficient strong
  updates. In: Proceedings of the 38th Annual ACM SIGPLAN-SIGACT Symposium on
  Principles of Programming Languages. p. 3–16. POPL '11, Association for
  Computing Machinery, New York, NY, USA (2011). \doi{10.1145/1926385.1926389},
  \url{https://doi.org/10.1145/1926385.1926389}

\bibitem{CRustS}
Ling, M., Yu, Y., Wu, H., Wang, Y., Cordy, J.R., Hassan, A.E.: In rust we trust
  - {A} transpiler from unsafe {C} to safer rust. In: 44th {IEEE/ACM}
  International Conference on Software Engineering: Companion Proceedings,
  {ICSE} Companion 2022, Pittsburgh, PA, USA, May 22-24, 2022. pp. 354--355.
  {ACM/IEEE} (2022). \doi{10.1145/3510454.3528640},
  \url{https://doi.org/10.1145/3510454.3528640}

\bibitem{10.1145/3527322}
Machiry, A., Kastner, J., McCutchen, M., Eline, A., Headley, K., Hicks, M.: C
  to checked c by 3c. Proc. ACM Program. Lang.  \textbf{6}(OOPSLA1) (apr 2022).
  \doi{10.1145/3527322}, \url{https://doi.org/10.1145/3527322}

\bibitem{matsakis2014rust}
Matsakis, N.D., Klock, F.S.: The rust language. In: Proceedings of the 2014 ACM
  SIGAda Annual Conference on High Integrity Language Technology. p. 103–104.
  HILT '14, Association for Computing Machinery, New York, NY, USA (2014).
  \doi{10.1145/2663171.2663188}, \url{https://doi.org/10.1145/2663171.2663188}

\bibitem{10.1145/3519939.3523704}
Matsushita, Y., Denis, X., Jourdan, J.H., Dreyer, D.: Rusthornbelt: A semantic
  foundation for functional verification of rust programs with unsafe code. In:
  Proceedings of the 43rd ACM SIGPLAN International Conference on Programming
  Language Design and Implementation. p. 841–856. PLDI 2022, Association for
  Computing Machinery, New York, NY, USA (2022). \doi{10.1145/3519939.3523704},
  \url{https://doi.org/10.1145/3519939.3523704}

\bibitem{pldi09}
Ravitch, T., Jackson, S., Aderhold, E., Liblit, B.: Automatic generation of
  library bindings using static analysis. In: Hind, M., Diwan, A. (eds.)
  Proceedings of the 2009 {ACM} {SIGPLAN} Conference on Programming Language
  Design and Implementation, {PLDI} 2009, Dublin, Ireland, June 15-21, 2009.
  pp. 352--362. {ACM} (2009). \doi{10.1145/1542476.1542516},
  \url{https://doi.org/10.1145/1542476.1542516}

\bibitem{10.1145/1122971.1122992}
Vafeiadis, V., Herlihy, M., Hoare, T., Shapiro, M.: Proving correctness of
  highly-concurrent linearisable objects. In: Proceedings of the Eleventh ACM
  SIGPLAN Symposium on Principles and Practice of Parallel Programming. p.
  129–136. PPoPP '06, Association for Computing Machinery, New York, NY, USA
  (2006). \doi{10.1145/1122971.1122992},
  \url{https://doi.org/10.1145/1122971.1122992}

\bibitem{10.1145/3485522}
Wolff, F., B\'{\i}l\'{y}, A., Matheja, C., M\"{u}ller, P., Summers, A.J.:
  Modular specification and verification of closures in rust. Proc. ACM
  Program. Lang.  \textbf{5}(OOPSLA) (oct 2021). \doi{10.1145/3485522},
  \url{https://doi.org/10.1145/3485522}

\bibitem{ZHAO2008213}
Zhao, T., Baker, J., Hunt, J., Noble, J., Vitek, J.: Implicit ownership types
  for memory management. Science of Computer Programming  \textbf{71}(3),
  213--241 (2008). \doi{https://doi.org/10.1016/j.scico.2008.04.001},
  \url{https://www.sciencedirect.com/science/article/pii/S0167642308000300}

\end{thebibliography}
\appendix
\section{Ownership constraints} \label{appendix:ownership}
The rules for the ownership monotonicity property, function body and function call, as well as
selected rules for modelling C library functions can be found in
\autoref{fig:selected-ownership-rules}.

{\bf MONOTONE.} When a new ownership variable is generated, \name generates a set of constraints meant to enforce the ownership monotonicity property as depicted by rule MONOTONE
in~\autoref{fig:selected-ownership-rules}.
In the rule, $V$ stands for the set of constraint variables, $C$ stands for the set of constraints, and $\Sigma$ stands for function signatures in terms of associated ownership variables.
According to the rule, if $p$ and $q$ are both access paths with base variable $v$ such that $p$ is a prefix of $q$, then the ownership of $p$ is
higher or equal to the ownership of $q$. For instance, \name generates the following for~\autoref{fig:llist-constr-rust}:
$\mathbb{O}_{\text{\lstinline{new_node}}}\ge\mathbb{O}_\text{\lstinline{(*new_node).next}}$.

For simpilicity, we only make $V$ or $\Sigma$ explicit in the rules if they are used. Also,
we assume that whenever a pointer is used, new variables are generated and MONOTONE
rule is applied. %

{\bf FREE.} When a pointer is passed to \lstinline{free}, we generate constraints that
assert that this pointer is owning prior to the call and non-owning after the call~\autoref{fig:selected-ownership-rules}.

{\bf FN-DECL.} The rule of generating constraints for function declarations is given as FN-DECL, where we consider a function ${\bf f}$
with output parameters ${\bf \vec{x}}$, normal parameters ${\bf \vec{y}}$, and function return ${\bf z}$.
This rule states that given current ownership variables $V$, constraints $C$ and function signature
$\Sigma$, the inference of the function declaration proceeds to generate ownership constraints
for statements in the function body with $V'$, $C'$, $\Sigma'$ updated accordingly.
For output parameter, we need to generate two sets of ownership variables, one on entry and one
on exit to represent its input/output ownership status, which we then constrain to be one.

{\bf FN-CALL.} The rule of calling a function is given as FN-CALL. As discussed in \autoref{sec:ownership-monotonicity},
we explicitly convert pointers \lstinline{p} to \lstinline{&mut (*p)} at output parameter positions.
Here we assume that all calling arguments of output parameters are in the form \lstinline{&mut p}.
The rule states that, for normal parameters, the ownership of arguments optionally may \emph{transfer}
to the parameters, as illustrated by the 3-variable constraints; for output parameter, the ownership 
of arguments gets \emph{borrowed} to the parameters: the entry/exit states of parameters are 
equated with pre/post states of arguments.

\begin{figure}
  \begin{mathpar}
    $\inferrule[MONOTONE]
    {p,q \in ap(v, 1, k)\\ \textit{is\_prefix}(p,q)\\ a = \textit{base\_var}(v)\\
    C'=C\cup\{\mathbb{O}_p {\geq} \mathbb{O}_{q}\}\\
    V'=V\cup\left\{\mathbb{O}_p\text{ new}\middle|p\in\textit{ap}\left(a, 1, k\right)\right\}
    }
    {V,C\vdash {\bf v}~\text{monotone}\Rightarrow V',C' }
    ~~~\inferrule[FREE]
    {
    C'=C\cup\{\mathbb{O}_v = 1\land \mathbb{O}_{v'} = 0\}}
    {C\vdash {\bf free(v)}\Rightarrow C'}
    \inferrule[FN-DECL]
      {\vec{v}\in\text{localVars}\left(\vec{\text{stmt}}\right)\\
      V' = V\cup\left\{\mathbb{O}_l\middle|l\in ap\left(\vec{x}_\text{entry}, 1, k\right)\cup ap\left(\vec{x}_\text{exit}, 1, k\right)\cup ap\left(\vec{y}, 1, k\right)\cup ap\left(z, 1, k\right)\right\}\\
      C' = C\cup\left\{\mathbb{O}_{\vec{x}}^\text{entry} = 1, \mathbb{O}_{\vec{x}}^\text{exit} = 1\right\}\cup\left\{\mathbb{O}_r = 0\middle| r\in ap\left(\vec{v}_\text{exit}, 1, k\right)\right\}\\
      \Sigma' = \Sigma\cup\{{\bf f}\left(\vec{{\bf x}};\vec{{\bf y}}\right):{\bf z}\}\\
      V',C',\Sigma'\vdash {\bf stmts} \Rightarrow V'',C''
      }
      {V,C,\Sigma\vdash{\bf f}\left(\vec{{\bf x}};\vec{{\bf y}}\right):{\bf z} \left\{\vec{\text{stmt}}\right\}\Rightarrow V'', C'',\Sigma'}
  \inferrule[FN-CALL]
    {
    f\left(\vec{x};\vec{y}\right):z \in \Sigma\\
    C' = C \cup \left\{\mathbb{O}_{r'} + \mathbb{O}_l = \mathbb{O}_r\middle| r \in ap\left(q, 1, k\right), l\in ap\left(y, 1, k\right), |r| - |q| = |l| - |y|\right\}\\
    \cup \left\{\mathbb{O}_{s} = \mathbb{O}_{n}, \mathbb{O}_{s'} = \mathbb{O}_{n'}\middle|s \in ap\left(p, 1, k\right), n\in ap\left(x, 2, k\right), |s| - |p| = |n| - |x| - 1\right\}\\
    \cup \left\{\mathbb{O}_t = 0 \land \mathbb{O}_{t'} = \mathbb{O}_m\middle|t\in ap\left(r, 1, k\right), m\in ap\left(z, 1, k\right), |t| - |r| = |m| - |z|\right\}}
    {C,\Sigma\vdash \text{\lstinline!let!}\ {\bf r}\ =\ {\bf f}\left(\text{\lstinline!&mut!}\ \vec{{\bf p}};\vec{{\bf q}}\right)\Rightarrow C'}
    $
  \end{mathpar}
  \caption{Selected ownership rules}
  \label{fig:selected-ownership-rules}
\end{figure}

\section{Proof of~\autoref{thm:monotonicity}}
\label{sec:proof-of-monotonicity}

  Let's consider a pointer assignment between $p$ and $q$.

  \noindent (i) If there is no ownership transfer, then the conclusion follows from the hypothesis.

  \noindent (ii) If ownership is transferred from $q$ to $p$, then $p'$ owns a new object after the assignment. For another pointer $l$ to own the same memory object after the assignment,
  $l$ must have been an alias of $p$ before the assignment. Let's now consider all possible access paths for $p$ and $l$:
 \begin{itemize}
 \item Both are accessed directly: $p = q$; $l$. In this case, $p'$ and $l'$ are no longer aliases after the assignment, meaning that they can't own the same object. 
 \item $p$ is accessed indirectly and $l$ directly: $*p=q$; $l$. If $*p$ aliases $l$, then, again, $*p'$ and $l'$ are no longer aliases after the assignment.
 \item Both are accessed indirectly: $*p=*q$; $*l$. In this case, we also need to consider the potential aliasing between $p$ and $l$.
   \begin{itemize}
   \item If $*p$ aliases $*l$, but $p$ does not alias $l$, then, again, $*p'$ and $*l'$ are no longer aliases after the assignment.  
   \item If $*p$ aliases $*l$ and $p$ aliases $l$, then, $*p'$ and $*l'$ are aliases after the assignment. Now,
   in order to check whether they can both be owning, let's look
   at the ownership constraints. We know that $*p'$ is owning, meaning that $\mathbb{O}_{*p'}=1$.
   By ownership monotonicity, we have that $\mathbb{O}_{p'}=1$. Given that $p'$ aliases $l'$, we know from the hypothesis that $l'$ can't own the same object, hence $\mathbb{O}_{l'}=0$.
   From ownership monotonicity we have $\mathbb{O}_{*l'} = 0$. Hence, $*l'$ can't be owning
   \end{itemize}
\item  Longer access paths follow the same proof as when $p$ is accessed indirectly above as it is sufficient to only look at the last indirection on the access path.
 \end{itemize}

\section{Proof of~\autoref{thm:transfer}}
\label{sec:proof-of-transfer}
\begin{proof}
  Let $a = \textit{base\_var}(v)$, $b = \textit{base\_var}(w)$.
  Suppose that $p\in\textit{ap}\left(a, |v|, k\right)$, $q\in\textit{ap}\left(b, |w|, k\right)$
  and additionally $|p| - |v| = |q| - |w|$. By rule ASSIGN, we have
  $\mathbb{O}_p = 0\land \mathbb{O}_{p'} + \mathbb{O}_{q'} = \mathbb{O}_{q}$. By rule MONOTONICITY,
  we have $\mathbb{O}_{w'} \ge \mathbb{O}_{q'}$. By the hypothesis, ownership transfers from
  $w$ to $v$, hence $\mathbb{O}_{w'} = 0$, which implies $\mathbb{O}_{q'} = 0$ and that
  $\mathbb{O}_{p'} = \mathbb{O}_q$. This means that if $q$ has ownership before the assignment, it will then be transferred to $p$.
  
\end{proof}

\section{Proof of \autoref{thm:ownership}} \label{appendix:thm-proof}
\begin{proof}

    Our proof consists of two parts:
    
    (1) The constraint solving is in NP. We show this by reducing it to SAT in polynomial time. In particular, each equality can be translated as follows:
    \begin{itemize}
    \item $x + y = z$ is translated to $(z \wedge x \wedge \neg y) \vee (z \wedge \neg x \wedge y) \vee (\neg z \wedge \neg x \wedge \neg y)$, whose CNF is
        $(z \vee y \vee \neg x) \wedge (x \vee  \neg y) \wedge (z \vee x \vee \neg y) \wedge (\neg y \vee \neg z) \wedge (x \vee \neg z) \wedge (x \vee y \vee \neg z)$.
    \item $x = 1$ is translated to $x$.
    \item $x = 0$ is translated to $\neg x$.      
    \item $x {\leq} y$ is translated to $(\neg x \wedge \neg y) \vee (x \wedge \neg y) \vee (x \wedge y)$  
    \end{itemize}

    (2) The constraint solving is NP-hard. We show this by reducing the EXACT-1-3-SAT problem to ownership constraint solving in polynomial time.
    EXACT-1-3-SAT is the problem of determining if there exists an interpretation that satisfies a given Boolean formula consisting of conjunctions of 3-literal clauses,
    with the extra restriction that exactly one literal is true per clause.
    
    We next describe how we construct the ownership constraint system.
    For each clause $l_1 \vee l_2 \vee l_3$, we generate three equalities in our constraint system: $l_1 + l_2 = not\_l_3$, $l_3 + not\_l_3 = l_4$ and $l_4 = 1$,
    where $not\_l_3$ and $l_4$ are fresh variables.
    Then, the original problem is satisfiable iff the ownership constraint system has a solution.
    Note that the only satisfiable configurations for a clause $l_1 \vee l_2 \vee l_3$ are (1, 0, 0), (0, 1, 0), (0, 0, 1), which are exactly the same as for $l_1 + l_2 = not\_l_3$, $l_3 + not\_l_3 = l_4$ and $l_4 = 1$, where, additionally, $not\_l_3 = \neg l_3$ and $l_4 = 1$. 
    Also, we know that each ownership variable has either value 0 or 1.
    \end{proof}

\end{document}